\documentclass[useAMS,usenatbib]{mn2e}

\usepackage{graphicx}
\usepackage{hyperref}
\usepackage{txfonts}
\usepackage{upgreek}
\usepackage{multirow}
\usepackage{longtable}

\title[Blazar Compton efficiencies]{Blazar Compton efficiencies: {\it Fermi}, external photons and the Sequence}

\author[Gupta et al.] {J.\,A.~Gupta, I.\,W.\,A.~Browne, M.\,W.~Peel \\
Jodrell Bank Centre for Astrophysics, Alan Turing Building, School of Physics and Astronomy, The University of Manchester, Oxford Road,\\
Manchester, M13 9PL}

\pagerange{\pageref{firstpage}--\pageref{lastpage}} \pubyear{2011}

\def\LaTeX{L\kern-.36em\raise.3ex\hbox{a}\kern-.15em  T\kern-.1667em\lower.7ex\hbox{E}\kern-.125emX}

\newcommand{\arxivpreprint}[1]{preprint (\href{http://arxiv.org/abs/#1}{arXiv:#1})}

\newcommand{\Fermi}{{\it Fermi}}

\begin{document}

\label{firstpage}

\maketitle

\begin{abstract}
The \Fermi-LAT survey provides a large sample of blazars selected on
the strength of their inverse Compton emission. We cross-correlate the
first \Fermi-LAT catalogue with the CRATES radio catalogue and use
this sample to investigate whether blazar gamma-ray luminosities are
influenced by the availability of external photons to be
up-scattered. Using the 8.4\,GHz flux densities of their compact radio
cores as a proxy for their jet power, we calculate their Compton
Efficiency parameters, which measure the ability of jets to convert
power in the form of ultra-relativistic electrons into Compton
gamma-rays. We find no clear differences in Compton efficiencies
between BL Lac objects and FSRQs and no anti-correlation between
Compton efficiency and synchrotron peak frequency. This suggests that
the scattering of external photons is energetically unimportant
compared to the synchrotron self-Compton process. These results
contradict the predictions of the blazar sequence.
\end{abstract}


\section{Introduction}
Some active galaxies produce relativistic jets from their nuclei and,
when the angle to the line of sight is small, the resulting Doppler
boosting is such that the non-thermal emission from the jets can
dominate the observed luminosity from radio through to gamma-ray
wavelengths. Collectively, such objects are labelled ``blazars''. At
radio, X-ray and gamma-ray wavelengths these are amongst the brightest
objects in the extragalactic sky (for a review of radio-loud active
galactic nuclei see \citealt{1995Urry}).

Blazars can be split into two types: flat spectrum radio quasars
(FSRQs) and BL Lac objects, though the boundaries between BL Lacs and
FSRQs can sometimes be a bit blurred
\citep[e.g.][]{1995Vermeulen}. The main observational difference
between the two types of blazar lies in their optical spectra; FSRQs
show strong broad emission lines, with continuum emission usually (but
not invariably) dominated by thermal emission believed to be from an
accretion disk, while BL Lacs lack strong emission lines and have
smooth optical continuum spectra believed to be of non-thermal origin.
Therefore, BL Lacs and FSRQs have large differences in their thermal
to non-thermal luminosity ratios. As a consequence, there is a large
variation in the availability of photons originating from regions
external from the jet that can be Compton scattered.

The spectral energy distributions (SEDs) of blazars are characterised
by emission produced by two different mechanisms, which results in a
distinct two-peak shape in log space. At longer wavelengths, blazar
non-thermal spectra are dominated by synchrotron emission from
ultra-relativistic electrons spiralling in the jet magnetic field. At
the shorter wavelengths the dominant emission process is thought to be
inverse Compton scattering of low-energy photons by the same
synchrotron electrons \citep{1981Konigl,1993Dermer,1998Fossati}. Both
emission mechanisms will reduce the energy of the relativistic
electrons. The highest energy electrons will radiate away most
of their energy either by the synchrotron or inverse Compton
mechanisms, depending upon which has the greater energy density: the
magnetic field in the jet or the low-energy photons. In the compact
regions at the base of the jets the inverse Compton mechanism will
dominate and this is where the gamma-ray emission from blazars that
has been detected in abundance by the \Fermi~gamma-ray space
telescope is thought to originate.

The seed photons for the inverse Compton emission can be produced
internally within the jet, and/or be the external AGN photons. The
radiation produced is referred to as synchrotron self-Compton (SSC) and
external Compton respectively \citep[e.g.][]{2008Ghisellini}. 
It has been proposed that the relative strength of external Compton
emission compared to SSC emission can explain the
diversity in the spectral energy distributions of blazars, and can
give rise to the ``blazar sequence''
\citep{1998Fossati,1998Ghisellini,2008Ghisellini,2009Sambruna}.

\section{Blazar sequence}

For more than a decade there has been extensive discussion focused on
the blazar sequence and the relative importance of external and
internally-generated photons in the inverse Compton process
\citep[e.g.][]{2007Padovani}.  The blazar sequence was originally
presented in two parts: a phenomenological aspect concerning an
observed anti-correlation between synchrotron peak frequency and radio
luminosity {\citep{1998Fossati}, and a theoretical counterpart
primarily concerning the abundance of external photons that can be
inverse Compton scattered \citep{1998Ghisellini}. More recently the
sequence framework has been extended to incorporate black hole mass
and accretion rate \citep{2008Ghisellini} but we will focus
exclusively on the role of external photons here.

The theoretical argument concerning external photons is that there
will be a cut-off in the energy spectrum of the synchrotron electrons,
and hence a frequency-dependent cut-off in the synchrotron
emission. The addition of external photons to scatter the most
energetic electrons should therefore decrease the frequency of the
cut-off. The presence of AGN disk and/or broad line emission in FSRQs
indicates that there are copious external photons available for
scattering. However, BL Lac spectra show no evidence for an external
source of photons associated with AGN emission. The primary
contention of the blazar sequence is that the more external photons
there are, the lower the expected peak frequencies of a blazar
SED. This expected trend is seen between FSRQs and BL Lacs. This has
been interpreted in different ways; either as strong observational
evidence supporting the blazar sequence \citep[e.g.][]{1998Fossati,
2008Ghisellini} or as a clear indication that BL Lacs and FSRQs
comprise physically distinct populations of objects
(e.g. \citealp{2005Anton}; also see \citealp{2007Padovani} for a
review). FSRQs generally have higher radio luminosities than BL Lacs,
thus a connection can be made between the theoretical argument and the
apparent anti-correlation between radio luminosity and synchrotron
peak frequency.

Though the phenomenological aspect of blazar sequence may be
controversial, there is little doubt that external Compton photons are
required to explain some of the individual SEDs in FSRQs, for example
in 3C454.3 \citep{2010Bonnoli}. 

It is important to remember that, in the active galaxies that produce
relativistic jets, only a small proportion of the bulk kinetic energy
in the outflow is converted into ultra-relativistic electrons. These
electrons in turn produce the electromagnetic signatures that enable
us to identify them. Though it is the product of the highest energy
electrons that dominate the electromagnetic spectrum, most of the
energy in the form of these synchrotron electrons is stored amongst
the lower energy electrons, not those with the highest energies. A key
issue, that is partially addressed by the blazar sequence discussion,
is what gives rise to the wide range of synchrotron peak frequencies
(which are indicative of the maximum energies of the synchrotron
electrons) in objects with apparently similar jet energies.

\section{Compton efficiency}\label{sec:comp_eff}

We introduce a simple concept of ``Compton efficiency'', which is a
measure of how efficiently power in the form of ultra-relativistic
electrons in a jet is converted by the inverse Compton process into
gamma-ray luminosity. In the following we will assume that all
gamma-rays are produced by inverse Compton scattering and ignore
non-leptonic processes \citep[e.g.][]{2003Mucke}.

\begin{figure}
\centering
\includegraphics[scale=0.15]{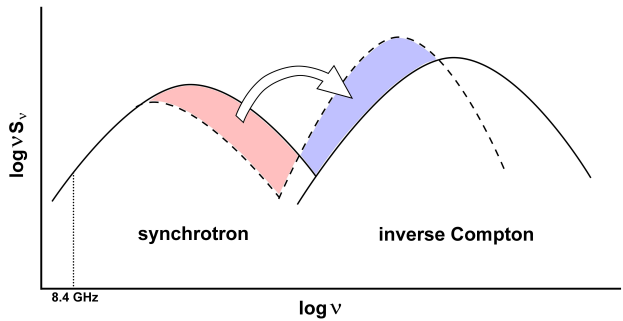}
\caption{Cartoon of blazar spectral energy distributions, illustrating
the key ideas of this paper. The solid SED represents a blazar with no
external photons. When a source of external photons is introduced,
energy from the synchrotron regime (left pink shaded area) is
transferred to the inverse Compton SED (right blue shaded area). The
resulting dashed SED has lower peak frequencies and higher inverse
Compton peak emission, but the low frequency radio emission is
unaffected.}
\label{fig:SED_cartoon}
\end{figure}

In the case of a jet whose energy losses are initially dominated by
SSC, as the external photon field is increased two things will happen:
the synchrotron peak frequency will decrease, and the integrated
gamma-ray luminosity will increase. However, the low frequency (radio)
synchrotron emission will remain unchanged. This is illustrated in
Figure \ref{fig:SED_cartoon} in terms of changes to the spectral
energy distribution caused by the presence of external
photons. Because the radio emission is immune to the energy loss
processes which only affect the highest energy electrons, we can use
the radio emission as a proxy for the energy content of the jet in the
form of relativistic electrons. Therefore to quantify the impact of
the external photon field, we use the ratio of the gamma-ray emission
at the inverse Compton (high frequency) peak of the SED to the compact
nuclear radio emission as a measure of the inverse Compton efficiency
$\epsilon$, i.e
\begin{equation}
\epsilon = \log_{10} \frac{(\nu S_{\nu})_\mathrm{IC,~peak}}{(\nu S_{\nu})_\mathrm{GHz}}.
\end{equation}

The choice of the best
measure of the gamma-ray emission required careful consideration. We
adopt $(\nu S_{\nu})_\mathrm{IC,~peak}$, where $\nu_\mathrm{IC,~peak}$
and $S_{\nu_\mathrm{IC,~peak}}$ are the frequency and the flux density
at which the inverse Compton SED peaks respectively. We do this
because we are trying to measure an energy efficiency and the peak in
the SED measures the maximum energy output. We cannot simply use the
measured gamma-ray flux density as a proxy because the different types
of blazar have a wide range of, and systematically different, inverse
Compton peak frequencies \citep{2009Abdo, 2010Abdob}. Implicit
in our definition of Compton efficiency is the assumption that both the
radio and gamma-ray emission have the same degree of relativistic
beaming and hence taking the ratio of observed luminosities cancels
the effects of different Doppler factors in different objects.

Our objective is to look for correlations of Compton efficiencies with
other observables. Most blazar samples are selected in the radio or
the X-ray and therefore contain objects selected by their synchrotron
emission, or a combination of synchrotron and inverse Compton
emission. However, a gamma-ray selected sample of blazars will be
selected purely by their inverse Compton emission. Thus, with such a
sample, we can safely compare the Compton efficiencies of different
types of object.

\section{\Fermi-LAT survey and other data} \label{sec:data}

\subsection{Input data}

The first \Fermi-LAT (Large Area Telescope) catalogue (1FGL) contains
1451 sources detected in the 0.1--100\,GeV energy range during the
first 11 months of science observations with
\Fermi~\citep{2010Abdob}. We use the online \Fermi-LAT point source
catalogue\footnote{\url{http://heasarc.gsfc.nasa.gov/W3Browse/fermi/fermilpsc.html}}
to cross-correlate these sources with those listed in the Combined
Radio All-Sky Targeted Eight GHz Survey (CRATES) compilation of
compact flat spectrum radio sources \citep{2007Healey}. The CRATES
catalogue excludes objects less than 10 degrees from the Galactic
plane to avoid contamination by Galactic radio sources. This gives us
a parent sample of 1043 \Fermi-LAT sources. We will work with two
subsamples: the FSRQs and the BL Lacs (as classified by 1FGL) of which
there are 271 and 288 respectively. When this sample is
cross-correlated with the CRATES catalogue, we obtain 224 FSRQs and
167 BL Lacs. The lower number of BL Lac correlations is likely due to
the different flux density limits of FSRQs and BL Lacs in the 1FGL
classification process, and the constant flux density limit within the
CRATES catalogue.

We choose to use the 8.4\,GHz flux density measurements from CRATES in
our Compton efficiency calculation because this frequency provides a
balance between probing the jet power and obtaining high enough
angular resolution to separate the core (jet) emission from any
extended emission. In particular, it combines high-resolution 8.4\,GHz
observations from the VLA in the northern hemisphere (predominantly
from the Cosmic Lens All-Sky Survey,
\citealp[CLASS;][]{2003Myers,2003Browne}) with similar observations
from ATCA in the southern hemisphere (predominantly from the AT20G
survey, \citealp{2004Ricci,2006Sadler}). However, a disadvantage of
the CRATES catalogue is that it lists each part of multi-component
sources separately. While the majority of multi-component sources have
an obvious dominant component that we take as the core, in some cases
the core position and flux density are not obvious. Since we use the
8.4\,GHz flux densities in our efficiency calculations, we have
examined the radio data for each source carefully to make sure that we
are using the correct value for the core radio emission. In two cases
(1FGL~J0856.6+2103 and 1FGL~J2149.7+0327) we have rejected the object
from our sample because we were uncertain about which of two radio
components were associated with the gamma-ray objects.  An issue with
the radio measurements is that most were made a decade before the
gamma-ray observations, so radio variability will introduce an extra
scatter in any measurement of Compton efficiency; we will discuss the
effects of this in Section 6.1.

\subsection{Gamma-ray peak}
As stated above, choosing the optimum gamma-ray quantity to use in the
definition of the Compton efficiency is important and we believe that
the value of $\nu S_{\nu}$ at the inverse Compton peak of the SED is
the best available. As such, we need to estimate the value of $\nu
S_{\nu}$ from the existing data. We use the recipe given by
\citet{2010Abdob} to estimate the inverse Compton peak frequency of an
SED given the measured photon index (which they also
provide). \citet{2010Abdob} also list gamma-ray flux densities in 6
bands in the 0.03--100\,GeV range; we use the 0.1--0.3, 0.3--1, 1--3
and 3--10\,GeV fluxes as the majority of our sources only have an
upper limit quoted in the lowest and highest energy bands.

Estimating the flux density at the peak frequency is easy when the
peak lies within the \Fermi\ energy range; we use two different
methods to calculate the peak flux density when this is not the
case. If there are sufficient flux density measurements, we use these
with the estimated peak frequency to fit a symmetric parabola. In
cases where there are insufficient flux densities, we make the
additional assumption that blazar SEDs have a characteristic width at
a given vertical distance from the peak. Using the 238 blazars that we
can fit a parabola to, we measure the width of the parabola at
$\log(\nu S_{\nu}) = \log(\nu S_{\nu})_\mathrm{IC,~peak} - 1$ and find
blazars have a median SED width $\Delta \log \nu_\mathrm{Hz} = 5.5$
(with a standard deviation of 1.39). This value is used to iteratively
fit a parabola to those sources with insufficient flux
measurements. The parabola is then used to calculate the gamma-ray
peak flux density.

\section{Results}
\subsection{2-colour plots}

\begin{figure}
\centering
\includegraphics[scale=0.32, angle=270]{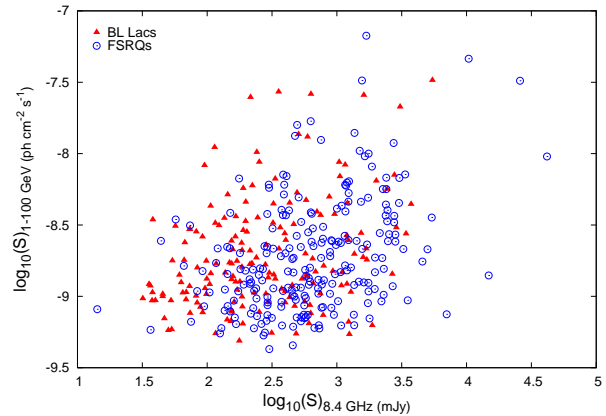}
\caption{CRATES 8.4\,GHz flux density against \Fermi\ 1--100\,GeV
flux showing a weak correlation. BL Lacs are shown by red filled triangles, and FSRQs are shown
by blue unfilled circles.}
\label{fig:flux_flux}
\end{figure}

\begin{figure}
\centering
\includegraphics[scale=0.32, angle=270]{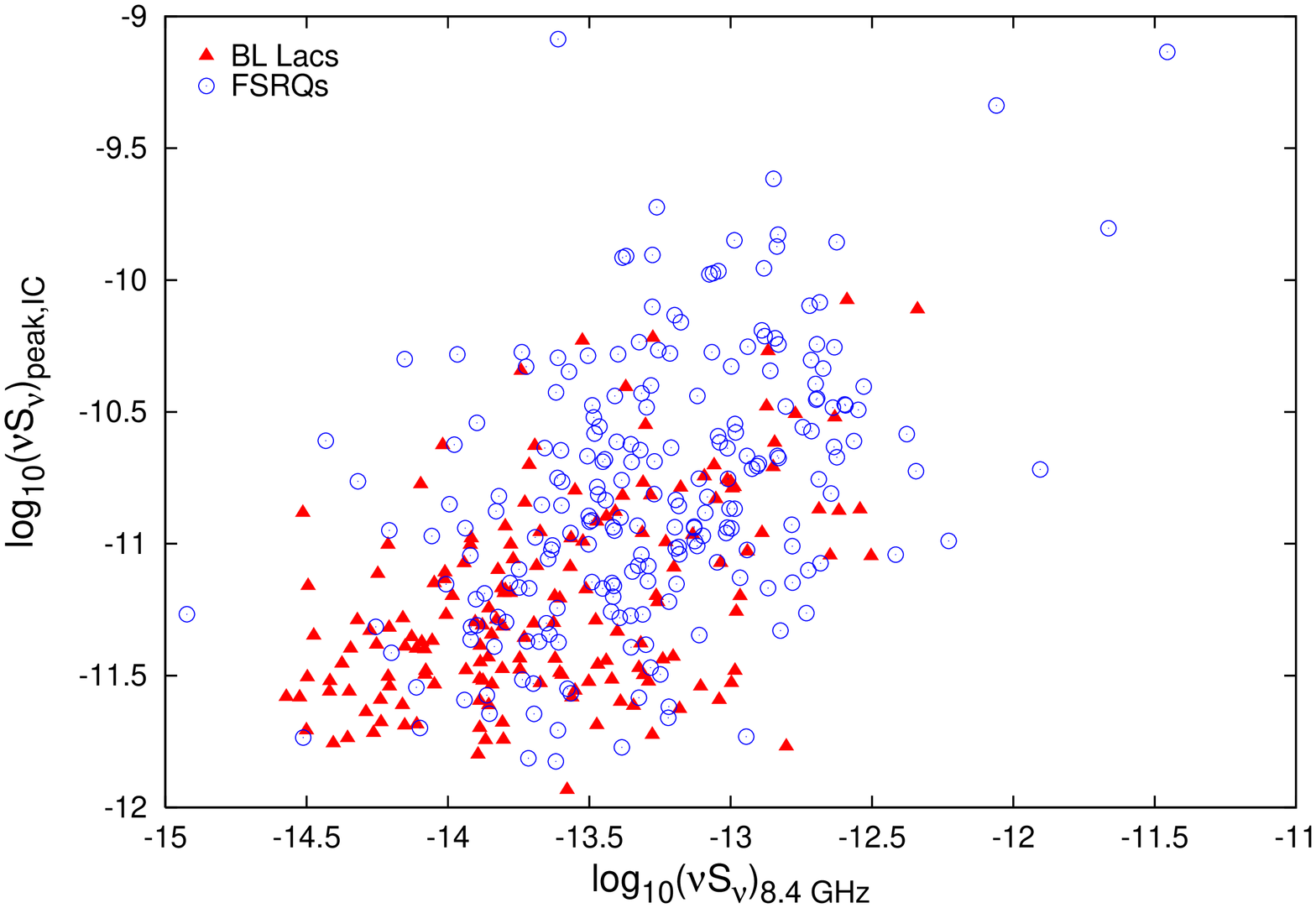}
\includegraphics[scale=0.32, angle=270]{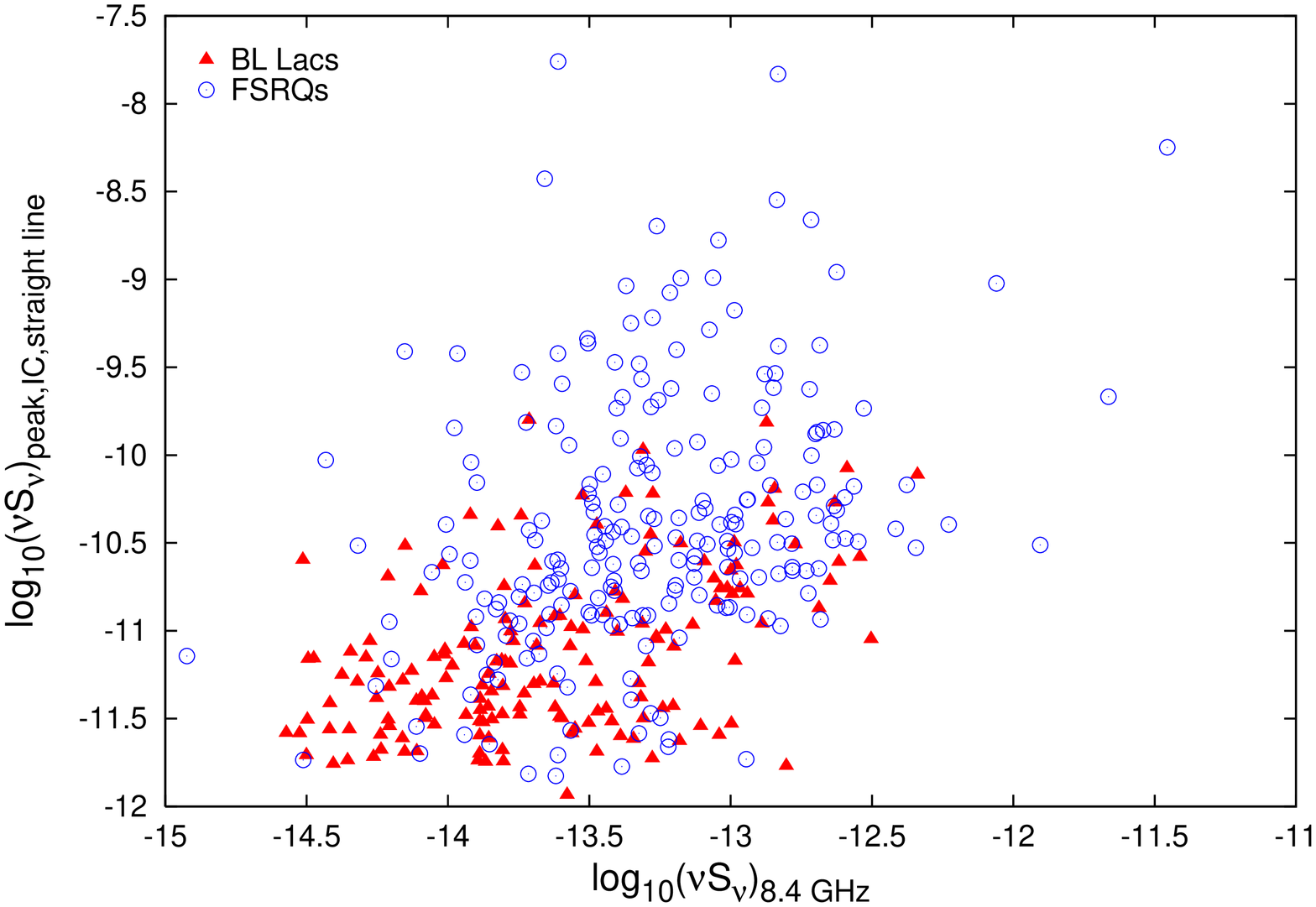}
\caption{{\bf Top:} 8.4\,GHz radio flux density (multiplied by $\nu$
for consistency) against $\nu S_{\nu}$ at the peak of the inverse
Compton SED regime showing a correlation. BL Lacs are shown by red
filled triangles, and FSRQs are shown by blue unfilled circles. {\bf
Bottom:} As top panel but with $\nu S_{\nu_{peak}}$ determined by a
linear fit.}
\label{fig:nuSnu}
\end{figure} 

In Figure \ref{fig:flux_flux} we show a plot of the 1--100\,GeV flux,
as detected by \Fermi-LAT, against the 8.4\,GHz flux density from
CRATES. Similar plots have been shown in e.g. \citet{2010Abdob,
2010Ghirlanda, 2011Peel}, but we repeat the plot here for
comparison. Two things are obvious from Figure \ref{fig:flux_flux}: BL
Lacs and FSRQs occupy overlapping but distinguishable areas and there
is a correlation between the gamma-ray and radio flux densities. From
this plot there is also an indication that for a given gamma-ray flux
BL Lacs have on average somewhat lower radio flux densities. This
would hint that BL Lacs might even have slightly higher Compton
efficiencies than FSRQs (in opposition to the expectations of the
original blazar sequence) but it could also arise from subtleties of
the identification process used by \citet{2010Abdob}.

The top panel of Figure \ref{fig:nuSnu} shows $\nu S_{\nu}$ at the
inverse Compton peak against the 8.4\,GHz flux density (multiplied by
$\nu$ for consistency). This plot shows many of the same qualitative
features as Figure \ref{fig:flux_flux}. There is still an indication
that, if anything, the BL Lac objects have on average less radio
luminosity for a given gamma-ray luminosity than the FSRQs. It is also
noticeable that the correlation between the quantities is stronger
(Spearman rank correlation coefficient $\rho = 0.534$ for Figure
\ref{fig:nuSnu}, top panel, compared to $\rho = 0.293$ in Figure
\ref{fig:flux_flux}) giving us encouragement that using the gamma-ray
peak $\nu S_{\nu}$ rather than the flux in a single band has
eliminated some of the bias due to the wide spread in peak
frequencies.

In order to see how sensitive the results are to the exact method used
to determine the gamma-ray peak frequencies, in the bottom panel of
Figure \ref{fig:nuSnu} we show the same quantities as the top panel
but with $\nu S_{\nu}$ at the peak determined by a straight line
extrapolation using the given photon index and flux densities. While
the two plots differ in detail, the overall trend and separation of
the two populations is still evident (Spearman rank correlation
coefficient $\rho = 0.513$). This implies that our results are not
very sensitive to the method used to determine the peak flux density.

\subsection{Compton efficiency distributions and correlations}

Our primary goal has been to compare the Compton efficiencies of BL
Lacs and FSRQs. There is no separation between BL Lacs and FSRQs
apparent in the top panel of Figure \ref{fig:cd} where we plot
histograms of the Compton efficiencies.  Application of the K-S test
shows that the probability that the distributions are drawn from the
same parent population is 38 per cent. Furthermore, it is clear that
the Compton efficiencies of FSRQs are certainly not higher than those
of BL Lacs as one might expect if the original blazar sequence
hypothesis were true; the median Compton efficiencies for FSRQs and BL
Lacs are $\epsilon = 2.44$ and $\epsilon = 2.47$ respectively. A
Wilcoxon rank sum test on the two distributions yields the result $P =
0.51$ i.e. the null hypothesis that the medians are equal cannot be
rejected at the 5 per cent level. However, we also find evidence that
this result could be influenced by selection effects (see Section
\ref{sec:sel_eff}).

\begin{figure}
\centering
\includegraphics[scale=0.32, angle=270]{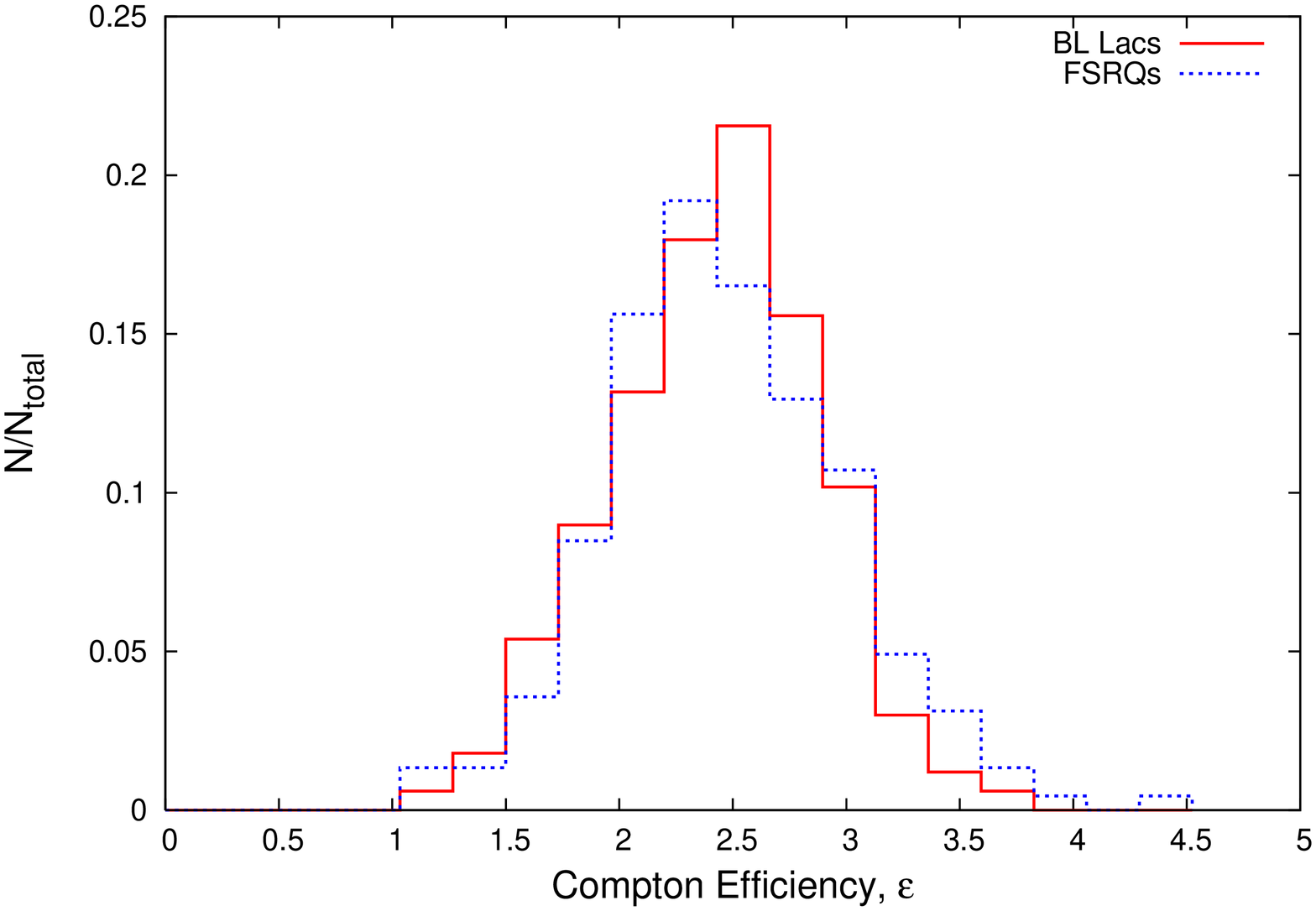}
\includegraphics[scale=0.32, angle=270]{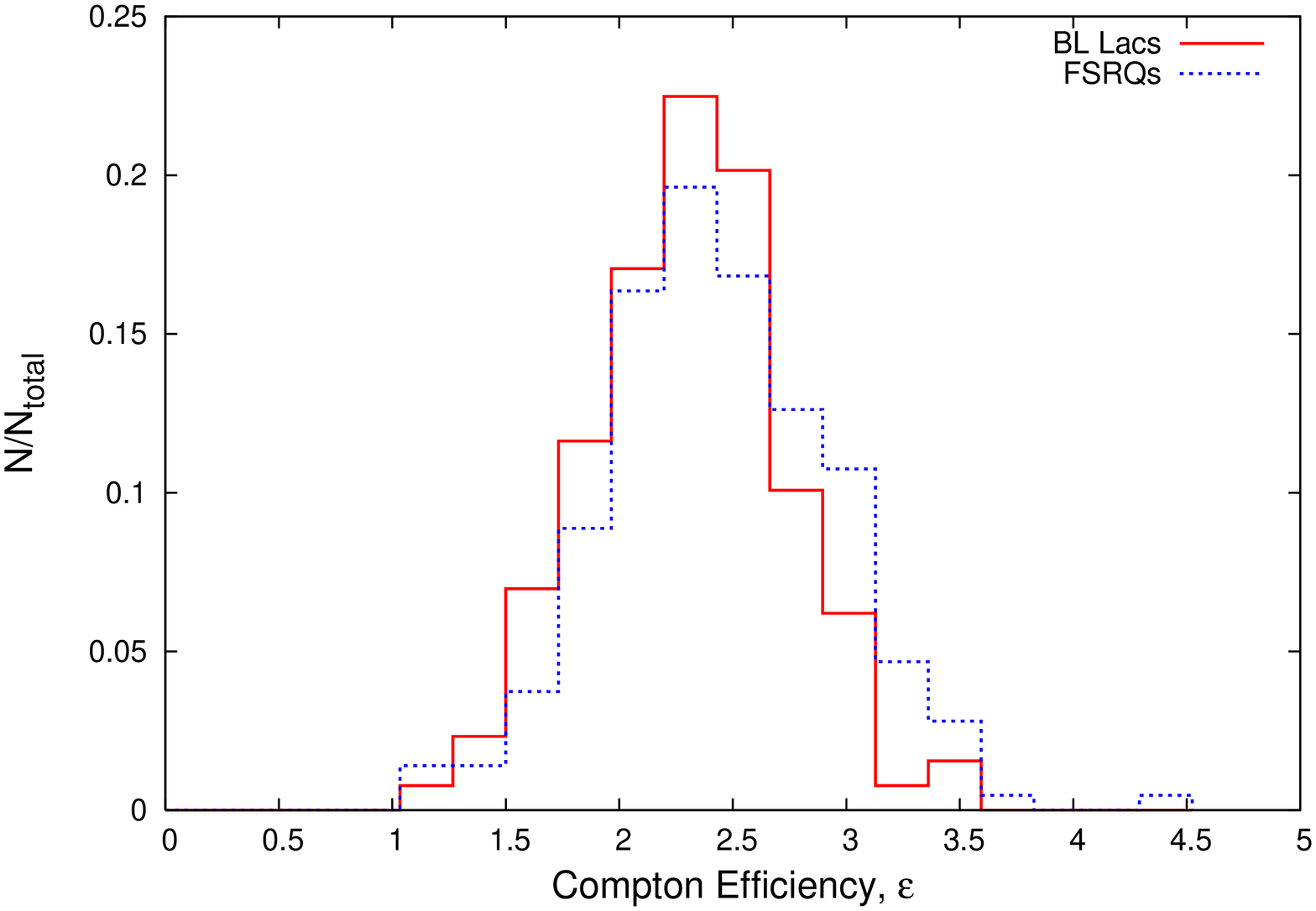}
\caption{{\bf Top:} Histogram showing the distribution of Compton
efficiencies for the two populations of blazars. BL Lacs are shown in
red (solid) and FSRQs are shown in blue (dashed). {\bf Bottom:} As top
panel but for sources with over 100~mJy of radio flux density.}
\label{fig:cd}
\end{figure}

In Figure \ref{fig:CE_syncp} we plot Compton efficiency against the
synchrotron peak frequency (for the subset of our objects that have a
synchrotron peak frequency provided by \citealp{2010Abdoa}). As stated
in Section \ref{sec:comp_eff}, in the simplest jet scenario, an
increase in external photons would decrease the synchrotron peak
frequency and produce a corresponding increase in gamma-ray emission,
but leave the radio unaffected. Thus an anti-correlation between
Compton efficiency and synchrotron peak frequency would be expected in
the blazar sequence framework. This is certainly not what is
observed. For the FSRQs, in all of which there is observational
evidence for a copious supply of external photons, there is no
indication of the expected anti-correlation, (Spearman rank test
coefficient $\rho = 0.088$).  Perhaps more significantly, for BL Lacs
there is a positive correlation (Spearman rank test coefficient $\rho
= 0.532$). Since BL Lacs are not so subject to the complications
arising from the presence of a strong optical AGN we suggest that the
observed correlation for these objects is a clear clue as to the
physical processes occurring in SSC jets.

\begin{figure}
\centering
\includegraphics[scale=0.32, angle=270]{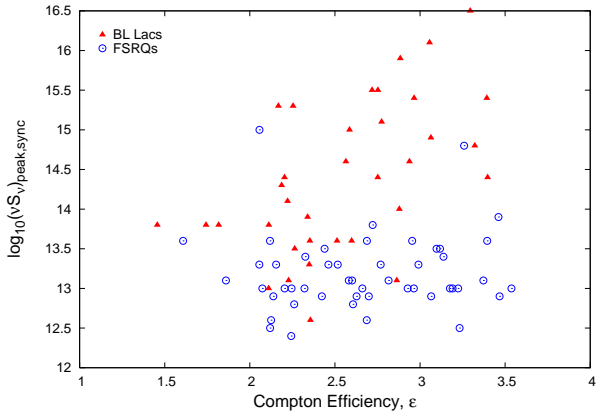}
\caption{Synchrotron peak frequency against Compton efficiency. BL
Lacs are shown by red filled triangles, and FSRQs are shown by blue
unfilled circles. A positive correlation is seen for BL Lacs while
there is no correlation for FSRQs.}
\label{fig:CE_syncp}
\end{figure}

\section{Discussion}\label{sec:disc}

We have tried to address a fundamental question concerning blazar
emission: given a certain amount of jet power, does the fraction of
that power that is converted into inverse Compton emission depend on
having visible AGN as a source of external photons?  The main result
of our analysis is the lack of any evidence that the Compton
efficiencies of FSRQs are any higher than those of BL Lacs. Also,
there is no evidence for an anti-correlation of synchrotron peak
frequency with Compton efficiency, which would be expected if
different concentrations of external photons were responsible for
shifting the peak frequency from blazar to blazar. Irrespective of the
details of any blazar sequence, it is somewhat surprising that the
availability of external photons to be scattered by the jet
synchrotron electrons does not appear to grossly affect the overall
energetics. The lack of separation between BL Lacs and FSRQs supports
the view that the processes within the jets are somehow immune from
outside environmental influences.

\subsection{Selection effects}\label{sec:sel_eff}

Before further discussion of the astrophysics we look at whether this
null result could simply be a consequence of selection effects. We
consider the following possibilities:

\begin{enumerate}

\item {\it Incomplete samples.} Of the 1043 1FGL sources at $|b| >
10^{\circ}$, 373 sources do not have associations with objects known
at other wavelengths. Many of these could be blazars contained in the
CRATES catalogue but not yet identified. If the identification process
systematically excluded high Compton efficiency FSRQs and not high
efficiency BL Lacs, this could bias our results.

For a gamma-ray selected sample, higher Compton efficiency goes with
weaker radio emission. Since the identification process is complicated
and has not been performed by ourselves we do not attempt to model its
effects. We do, however, note that in Figure \ref{fig:flux_flux} the
ratio of FSRQs to BL Lacs appears to decrease with radio flux
density. For this reason we have repeated the previous analysis only
using radio sources stronger than 100~mJy. The results are shown in
the bottom panel of Figure \ref{fig:cd} from which it can be seen that
excluding weak radio sources reveals a difference in the Compton
efficiency distributions for BL Lacs and FSRQs; the K-S test gives a
probability of 2.2 per cent that the two distributions are drawn
from the same parent population. Furthermore, the median Compton
efficiencies are now $\epsilon = 2.33$ for BL Lacs and $\epsilon =
2.40$ for FSRQs; the Wilcoxon rank sum test rejects the null
hypothesis that the medians are the same, at the 5 per cent level ($P
= 0.012$). This indicates that our null result could be due to the
incompleteness of our sample (especially for FSRQs) at low radio flux
densities.

\item {\it Radio variability.} Most of the radio flux densities we
have used in the Compton efficiency calculations were made more than a
decade ago and thus radio variability will to first order increase the
dispersion on the efficiency numbers. However, variability on
timescales of years at 8.4\,GHz is rarely by more than a factor of
two. For example, \citet{2010Jackson} recently re-measured 8.4\,GHz
flux densities for $\sim$100 strong CRATES/CLASS sources and found
that $\leq$15 per cent had varied by more than a factor of two over
that time interval. This is small compared to the more than an order
of magnitude dispersion in Compton efficiencies and thus it would
appear that most of the dispersion in efficiencies is intrinsic and
not due to the fact that the radio and gamma-ray flux densities were
not measured coevally. While it might be useful to have
contemporaneous measurements, with the present limited numbers of
objects it would be unlikely to affect the conclusions.

\end{enumerate}

There is some evidence that selection effects arising from
incompleteness at low radio flux densities may be responsible for our
result that there is no separation between the Compton efficiency
distributions of BL Lacs and FSRQs. Hence further work to find radio
counterparts to the unidentified 1FGL sources is needed before final
conclusions can be drawn from the Compton efficiency
distributions. However, such selection effects do not apply our
conclusions from drawn from Figure \ref{fig:CE_syncp} where we are
looking for the predicted correlation within the FSRQ population
itself and fail to find it.

\subsection{Compton efficiencies and the blazar sequence}

The blazar sequence invokes the relative importance of a reservoir of
photons external to the jet and those internally generated to
account for the systematic differences in the SEDs of blazars. Another
assumption underlying the blazar sequence hypothesis is that the bulk
relativistic motion, particle acceleration and magnetic fields
(i.e. those properties that determine the low frequency synchrotron
emission) of all blazar jets have approximately the same distributions
of values irrespective of whether they are hosted by BL Lacs or FSRQs.
Our investigation tries to explore the consequences of such a
scenario. Given the assumptions above we would expect two things: 

\begin{enumerate}

\item We would expect the synchrotron peak frequencies and the Compton
peak frequencies for BL Lacs to be higher than those for FSRQs because
of external Compton scattering. This is what is observed and is the
the main theoretical justification for the blazar sequence.

\item We would expect a higher proportion of the energy in
ultra-relativistic particles to be converted into gamma-rays in the
FSRQs than the BL Lacs. The observable effect of this should be higher
Compton efficiencies in FSRQs than BL Lacs and an anti-correlation
between Compton efficiency and synchrotron peak frequency.

\end{enumerate}

Although there is an indication that the Compton efficiencies of FSRQs
are higher than for BL Lacs when taking selection effects into
account, we do not see the expected anti-correlation with synchrotron
peak frequency. The obvious conclusion is that some assumption
underlying the blazar sequence is invalid.  Either BL Lacs and FSRQs
represent physically distinct populations (as indicated in the bottom
panel of Figure \ref{fig:cd}) in which case our result that their
Compton efficiencies are indistinguishable has to be a coincidence,
or, external Compton is rarely energetically dominant.

\section{Summary and conclusions}

We have defined a Compton efficiency parameter that measures the
amount of energy in the jet that has been converted into inverse
Compton emission. It is a more direct measure of a fundamental
property of blazars than has previously been used. Based on the
assumptions of the blazar sequence, we would expect the Compton
efficiencies of FSRQs and BL Lacs to be different, and that an
anti-correlation would exist between Compton efficiency and
synchrotron peak frequency. We do not see these results in the full
1FGL \Fermi~data, although by excluding weak radio sources, we see an
indication that selection effects could play a role in our results. On
the other hand, based on the same assumptions we would also expect to
see an anti-correlation between Compton efficiency and synchrotron
peak frequency. We do not see the expected anti-correlation; this is a
more robust conclusion because it is independent of possible selection
biases that may affect FSRQs and BL Lacs differently. Moreover,
amongst BL Lacs alone there is a positive correlation, which we
suspect is a useful clue about the physics of blazar jets.

The obvious conclusion is that the availability of a source of
external photons to be scattered to high energies does not have a
dominant effect on the overall gamma-ray luminosities of
blazars. However, it is conceivable that jets with multi-zone emission
regions, possibly having different Doppler boosting factors, could
account for our results. In the future larger and more completely
identified samples of \Fermi-selected sources will become available
that should significantly improve the statistics and reduce selection
biases.

\section*{Acknowledgements} 
 
We thank Neal Jackson and Sonia Anton for useful discussions. JG
acknowledges the support of an STFC studentship. This research has
made use of the NASA/IPAC Extragalactic Database (NED) which is
operated by the Jet Propulsion Laboratory, California Institute of
Technology, under contract with the National Aeronautics and Space
Administration.

\label{lastpage}
\end{document}